\documentclass[smallextended]{svjour3}       %
\smartqed  
\usepackage{graphicx}
\usepackage{amsmath}
\usepackage{latexsym}
\usepackage{amssymb}
\usepackage{graphicx}
\usepackage[colorlinks=true, citecolor=blue, urlcolor=blue]{hyperref}
\usepackage{float}
\usepackage{amsfonts}
\usepackage{textcomp}
\usepackage{caption}
\usepackage{lipsum}

\begin{document}

\title{Quantum Nash equilibrium in the thermodynamic limit
}


\author{  Shubhayan Sarkar        \and
       Colin Benjamin 
}
\institute{ Shubhayan Sarkar \at
              School of Physical Sciences, National Institute of Science Education \& Research, HBNI, Jatni-752050, India \\
           \and
          Colin Benjamin\at
         School of Physical Sciences, National Institute of Science Education \& Research, HBNI, Jatni-752050, India \\   
 \email{colin.nano@gmail.com}}
\date{Received: date / Accepted: date}
\maketitle
\begin{abstract}
The quantum Nash equilibrium in the thermodynamic limit is studied for games like quantum Prisoner's dilemma and quantum game of Chicken. A phase transition is seen in both games as function of the entanglement in the game. We observe that for maximal entanglement irrespective of the classical payoffs, majority of players choose quantum strategy over defect in the thermodynamic limit.
\keywords{Quantum games \and Hawk-Dove game \and Nash equilibrium}
\PACS{02.50.Le,01.80.+b,03.65.Ud,03.67.Ac}
\end{abstract}
\section{Introduction}
Quantum game theory is an important extension of classical game theory to the quantum regime. The classical games might be quantized by superposing initial states, entanglement between players or superposition of strategies, for a brief account see~\cite{12}. The outcomes of a quantum game is well known for two player case however, we want to investigate the scenario when the number of players goes to infinity, i.e., the thermodynamic limit. In recent times, there have been attempts to extend the two player classical games to the thermodynamic limit by connecting it to the Ising model~\cite{9,3,10}. We do a similar analysis and connect two player quantum games to the 1D Ising model in the thermodynamic limit to figure out the strategy chosen by majority of the population and try to predict the Nash equilibrium when the choices are entangled. 

We approach the problem similar to as done in Refs.~\cite{3,10}, where players are equivalent to sites and spins at each site represent the strategies of the players. From a game theoretic perspective, Magnetisation is defined as the difference between the average number of players opting for strategy $s_1$ over $s_2$. We first understand the connections between two player games and two spin Ising Hamiltonian and then extend it to the thermodynamic limit. We then quantize the Prisoner's dilemma and game of Chicken using the Eisert's scheme~\cite{2}. We find the payoffs corresponding to the Quantum strategy (Quantum strategy will be explicitly defined in section III). We then extend these games to the thermodynamic limit by considering classical strategies (cooperation or defection) against the Quantum strategy. We see that for all the games considered, Quantum and Cooperation are equiprobable strategies. However, we note an intereseting feature that when the entanglement between the players becomes maximum, then the majority of players would always choose the Quantum strategy and won't defect irrespective of the game's payoffs. We also see how quantum Nash equilibrium changes when the entanglement in the system is varied.

This paper is organised as follows- In section IA, we review the 1-D Ising model and the connection of two spin Ising Hamiltonian to the payoffs of a general two player classical game as in Ref. \cite{1}. We then extend it to the thermodynamic limit using the approach of Ref. \cite{3,10} for a classical game. Next, we quantize the Prisoner's dilemma game using Eisert's scheme \cite{2} and then make the connection for the thermodynamic limit of quantum games, where entanglement plays a non-trivial role. We then calculate the Nash equilibrium in the thermodynamic limit for the quantum Prisoner's dilemma. In section III, we again quantize the game of Chicken using Eisert's scheme and predict the Nash equilibrium in the thermodynamic limit. We plot the probability of choosing one strategy over other versus the entangling factor $\gamma$, and observe the phase transitions as we vary the parameters. Interestingly we find that in both games for maximal entanglement, the majority of individuals always opt for the quantum strategy. 
\subsection{Classical game theory and 1D Ising model}\label{sec2}
In the 1D Ising model \cite{6}, the spins are put on a 1D line and are in either of the two states $+1$ ($\uparrow$)  or $-1$ ($\downarrow$). The interaction is restricted between nearest neighbors only. The Hamiltonian of the 1D Ising  system is given as-
\begin{equation}\label{eq10}
H=-J\sum^N_{k=1}\sigma_k\sigma_{k+1}-h\sum^N_{k=1}\sigma_k,
\end{equation}
where $J$ denotes the spin-spin coupling, $h$ is the external magnetic field and the spins are denoted by $\sigma$'s. Using the above Hamiltonian Eq.~(\ref{eq10}), the Magnetisation can be derived\cite{6} as- 
\begin{equation}\label{eq8}
m=\frac{\sinh(\beta h)}{\sqrt{\sinh^2(\beta h)+e^{-4\beta J}}}.
\end{equation}
A two spin 1D Ising Hamiltonian and the payoff matrix for a two player game can be mapped as was shown in Ref.~\cite{1,3}. A general two player game has a payoff matrix given by-
\begin{equation}\label{eq9}
U=\left(\begin{array}{c|cc} & s_1 & s_2 \\\hline s_1 & a,a' & b,b' \\  s_2 & c,c' & d,d'\end{array}\right),
\end{equation} 
where $U(s_i,s_j)$ is the payoff function and $a, b, c, d$ are the row player's payoffs while $a', b', c', d'$ are the column player's payoffs. The strategies adopted by the two players are denoted as $s_1$ and $s_2$. To extend a two player game to a N-player game, i.e., the thermodynamic limit we proceed by first defining the two player Ising game matrix corresponding to a two player game, as in Eq.~\ref{eq9}. For a full derivation of Ising game matrix from a two spin Ising model, see Refs.~\cite{3,10,1} It is then straightforward to calculate the magnetization of the N player game, i.e., the difference between number of players opting for strategy $s_1$ over $s_2$ using Eq.~\ref{eq8}. 

To map the two player game to the  Ising game matrix we proceed as follows- a factor $\lambda$ is added to the $s_1$ column and $\mu$ to the $s_2$ column in Eq.~\ref{eq9}. Thus, we have-    
\begin{equation}\label{eq2}
U=\left(\begin{array}{c|cc} & s_1 & s_2 \\\hline s_1 & a+\lambda & b+\mu \\  s_2 & c+\lambda & d+\mu\end{array}\right).
\end{equation} 
As shown in Ref.~\cite{1,3}, the Nash equilibrium remains invariant under such a change to the payoffs. {To show this we used the fixed point analysis of game theory. Since the Nash equilibrium corresponds to a fixed point, we show that the fixed point corresponding to the game Eq.~(3) and the transformed game Eq.~(4) are the same, see appendix of Ref.~[3] for a detailed explanation of the invariance of Nash equilibrium fixed point.} The Ising game matrix~\cite{3,10,1} is defined as-
\begin{equation}\label{eq13-}
\left(\begin{array}{c|cc}  & s_2=+1 & s_2=-1  \\\hline s_1=+1 & J+h & -J+h \\s_1=-1 & -J-h & J-h\end{array}\right).
\end{equation} 
where $s_1,\ s_2$ denotes the spin at a particular site. The players in game theory are represented by the site and the strategies are represented by spins in the Ising model (see Refs.~\cite{3,1} for a detailed derivation). Choosing the transformations as $\lambda=-\frac{a+c}{2}$ and $\mu=-\frac{b+d}{2}$ in Eq.~(\ref{eq2}), the elements of Ising game matrix Eq.~(\ref{eq13-}) can be mapped directly to the transformed payoff matrix Eq.~(\ref{eq9}). Thus, we get the values of $J,h$ which define the N-player game in terms of the row player payoff's of the two player game as-
\begin{eqnarray*}
J=\frac{a-c+d-b}{4},\ h=\frac{a-c+b-d}{4}.
\end{eqnarray*} 
The Magnetization which gives the difference between the average number of players opting for strategy $s_1$ over $s_2$, from Eq.~(\ref{eq8}) can be written in terms of the payoff matrix elements Eq.~(\ref{eq9}) as-
\begin{equation}\label{eq13}
m=\frac{\sinh(\beta \frac{a-c+b-d}{4})}{\sqrt{\sinh^2(\beta \frac{a-c+b-d}{4})+e^{-\beta (a-c+d-b)}}}.
\end{equation} 
Eq.~(\ref{eq13}) defines the connection between the payoffs from a classical two player game with the magnetization of the N-player game, i.e., the thermodynamic limit. {To summarize the methodology, we first map the general two-player game payoff matrix Eq.~(3) to the 2-site Ising game matrix Eq.~(5) by adding payoff factors-$\lambda$ and $\mu$ to the columns of Eq.(3). The addition of factors helps to make a one-to-one correspondence between two site Ising game matrix and payoffs of a two player game. Further, under such transformations the Nash equilibrium remains invariant, see Appendix of Ref.~[3]. Equating the Ising game matrix to the transformed payoff matrix Eq.~(4) we find the parameters of 1-D Ising model ($J$ and $h$) in terms of game payoffs. The 2-site Ising game matrix is just a subset of the N-site Ising model, which for $N\rightarrow \infty$ gives the thermodynamic limit (from statistical physics). Thus, the magnetization of the Ising model in the thermodynamic limit is now expressed in terms of the game payoffs which is effectively the difference in the fraction of players choosing one strategy over other. Thus, we can get the distribution of strategies in the thermodynamic limit by mapping the two-player payoff matrix to the 2-site Ising model.  An account of infinite player games has also been attempted in Ref.~[8] to study Nash equilibrium using a different approach, but unlike this work which  focuses on  quantum games, it is classical and further it does not deal with the question on how cooperation arises in the infinite player case.}

When temperature in Ising system increases, i.e., $\beta=\frac{1}{k_{B}T}$ decreases, the spins become more disordered. Similarly, decreasing $\beta$ in game theory relates to increasing the randomness in choices for individual players. Now to  connect  quantum game theory to the 1D Ising model so as to find the quantum Nash equilibrium in the thermodynamic limit we consider first the Prisoner's dilemma. We first quantize the classical two player Prisoner's dilemma incorporating entanglement and then model the mapping to a N-player quantum prisoners dilemma by taking recourse to the Ising game matrix as shown in Eqs.~\ref{eq13-} and then similarly calculating the magnetization of the N-player quantum prisoner's dilemma, see Eq.~\ref{eq13}.
\section{Prisoner's dilemma}
In the classical Prisoner’s dilemma game, police interrogate two suspects separately. Each suspect can either cooperate with the other and not admit the crime (C) or defect against the other(D) implicating him in the crime. The payoff matrix is constructed by taking the matrix elements from Eq.~(\ref{eq9}) as $a=r$, $d=p$, $b=s$ and $c=t$, with the condition $t>r>p>s$. The reward is given by $r$, temptation is $t$, $s$ is  sucker's payoff and the punishment is given by $p$. Thus, the classical payoff matrix is given by- 
\begin{equation}\label{eq15}
U=\left(\begin{array}{c|cc} & C & D \\\hline C & r,r & s,t \\  D & t,s & p,p\end{array}\right).
\end{equation} 
Independent of the strategy opted by the fellow suspect, one can always stay safe by defecting. Thus, the Nash equilibrium in classical Prisoner's dilemma is always defection.
\subsection{Quantum game theory and 1D Ising model}\label{sec2.1}
{The  classical Prisoner's dilemma game was quantized by Eisert, et. al.}, in Ref.~\cite{2}. We explain their procedure by representing the convicts as qubits and their strategies as the state of those qubits. The cooperation strategy is represented as $|0\rangle$ while defection is represented as $|1\rangle$. To choose a particular strategy the operator $O(\theta,\phi)$ is applied on the initial state where,
\begin{equation}\label{unitary}
O(\theta,\phi)=\left(\begin{array}{cc} e^{i\phi}\cos(\theta/2) & \sin(\theta/2) \\ -\sin(\theta/2) & e^{-i\phi}\cos(\theta/2) \end{array}\right).
\end{equation}
The operator $\hat{L}$  entangles the choices-
\begin{equation}
\hat{L}=\left(\begin{array}{cccc} \cos(\gamma/2)& 0 & 0 & i \sin(\gamma/2)  \\  0& \cos(\gamma/2) & -i \sin(\gamma/2) & 0 \\  0 & -i \sin(\gamma/2)  & \cos(\gamma/2)& 0 \\ i \sin(\gamma/2) & 0 & 0 & \cos(\gamma/2) \end{array}\right),\nonumber
\end{equation}
with $\gamma$ being a measure of the entanglement in the game. $\gamma=0$ implies no entanglement while $\gamma=\pi/2$ implies maximal entanglement. The initial state $|00\rangle$ on being acted by $\hat{L}$ gives- 
$|\psi_k\rangle=\ \cos(\gamma/2)|00\rangle+i\ \sin(\gamma/2)|11\rangle$, the subscript $k$ indicates the site index and the final state after the action of the disentangling operator $\hat{L}^{\dagger}$ and the unitary operators $O$'s representing the strategies is-
\begin{equation}
|\chi_k\rangle=\hat{L}^{\dagger}O(\theta_1,\phi_1)\otimes O(\theta_2,\phi_2)\hat{L}|00\rangle.
\end{equation}
The classical prisoner's dilemma payoffs are given by Eq.~(\ref{eq15}). The payoffs for qubits $A$ and $B$ are then calculated using the template of classical prisoner's dilemma as-
\begin{eqnarray}
\$_A&=&r|\langle00|\chi_k\rangle|^2+t|\langle10|\chi_k\rangle|^2+s|\langle01|\chi_k\rangle|^2+p|\langle11|\chi_k\rangle|^2,\nonumber\\
\$_B&=&r|\langle00|\chi_k\rangle|^2+s|\langle10|\chi_k\rangle|^2+t|\langle01|\chi_k\rangle|^2+p|\langle11|\chi_k\rangle|^2.\nonumber\\
\end{eqnarray}  
Inclusion of quantum strategy via $Q=i Z=O(0,\pi/2)$ leads to a new payoff matrix:
\begin{equation}\label{eq24}
U=\left(\begin{array}{c|ccc} & C & D & Q\\\hline C & r,r & s,t  & \alpha 1,\alpha 1\\  D & t,s & p,p & \alpha 2,\alpha 3 \\Q & \alpha 1,\alpha 1 & \alpha 3,\alpha 2 &  r,r\end{array}\right),
\end{equation}
where $C=I$ and $D=X$ and $\alpha 1=r\cos^2(\gamma)+p\sin^2(\gamma)$, $\alpha 2=s\cos^2(\gamma)+t\sin^2(\gamma)$, and $\alpha 3=t\cos^2(\gamma)+s\sin^2(\gamma)$. When $\gamma=\pi/2$,  $|\psi_k>$ is a maximally entangled state and  the payoff matrix becomes taking- $r=3$, $t=5$, $s=0$ and $p=1$:
\begin{equation}\label{3x3}
U=\left(\begin{array}{c|ccc} & C & D & Q\\\hline C & 3,3 & 0,5  & 1,1\\  D & 5,0 & 1,1 & 0,5 \\Q & 1,1 & 5,0 &  3,3\end{array}\right).
\end{equation} 
As we can clearly see that there are two optimal strategies, i.e., both players can choose to cooperate or the quantum strategy. However, the new Nash equilibrium is the quantum strategy. Thus, if the players choose the quantum strategy then they will be at equilibrium with the maximum benefit shared between both the players.

\begin{figure}
\includegraphics[width=.99\linewidth]{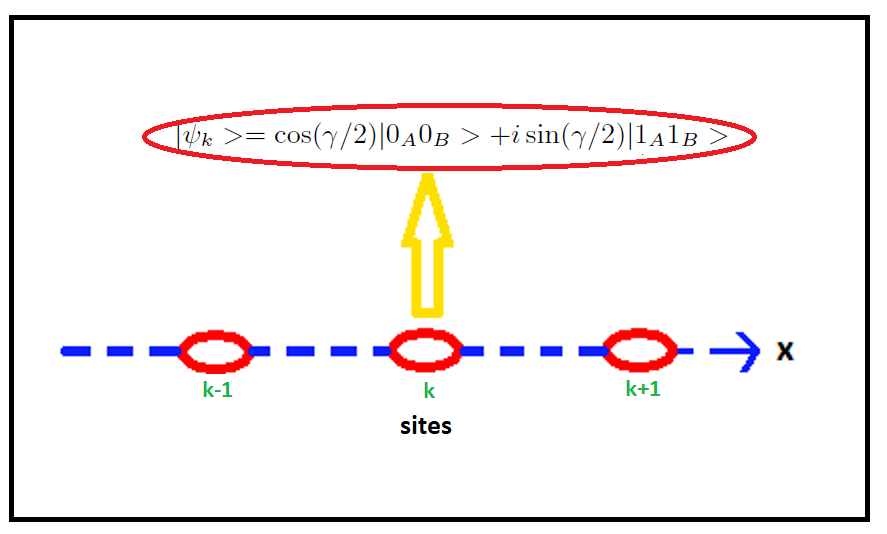}
  \caption{Representation of Ising model and its extension to quantum game theory. The sites  are represented by the site index $k$. The blue dashed line represents classical coupling $J$ between sites while in each site an entangled state is present on whose two qubits the quantum game is played.}
  \label{fig8:}
\end{figure}

Now to extend the above two player game to the infinite player limit or the thermodynamic limit, we proceed as follows.  Similar to the extension of classical games to the thermodynamic limit wherein the strategies or choices of the players are represented by the spins in the Ising model, herein each site plays the role of a spin in the classical Ising model, see Fig.~\ref{fig8:}. However, to incorporate entanglement each site contains an entangled pair and a two player quantum game is played at each site.  A site interacts via classical coupling factor $J$ with its neighboring site. $h$ the external magnetic field which tends to align the spins in a particular direction in the classical Ising model, herein plays the role of an external parameter which tends to make the sites behave similar to each other in the N-site quantum game. A schematic diagram is shown in Fig.~\ref{fig8:}.
To map the payoffs from the two player quantum prisoner's dilemma at a single site to the Ising game matrix~(\ref{eq13-}) we have each player in a game to have access to either classical or quantum strategy, see Eq.~(\ref{unitary}). We investigate the behavior of the N-site population for classical versus quantum strategy played at each site. The magnetization for the N-site population is then calculated as before from Eqs.~(\ref{eq13-},\ref{eq13}). Thus we break the payoff matrix~(\ref{3x3}) in to two separate $2\times2$ blocks: one for Quantum versus Cooperation and the other block for Quantum versus Defect.
\subsection{Quantum versus Cooperation}
As discussed in section~\ref{sec2.1} we calculate the magnetization in a scenario when Alice and Bob have access to only quantum ($Q=iZ$) and cooperation ($C=I$) strategy. The payoff matrix for the row player then from Eq.~(\ref{3x3}) is
\begin{equation}\label{QvC}
U=\left(\begin{array}{c|cc} & C & Q \\\hline C & r & r\cos^2(\gamma)+p\sin^2(\gamma) \\  Q & r\cos^2(\gamma)+p\sin^2(\gamma) & r\end{array}\right).
\end{equation}
We see that when $\gamma=0$, the quantum strategy reduces to the cooperation strategy in the quantum Prisoner's dilemma. As we can see from the payoff matrix, there are two Nash equilibriums- both the players can choose to cooperate or choose quantum. Transforming the matrix~(\ref{QvC}) using the method explained in section~\ref{sec2}, we get the transformations as-
$\lambda=-\frac{a+c}{2}=-\frac{r+r\cos^2(\gamma)+p\sin^2(\gamma)}{2}$ and $\mu=-\frac{b+d}{2}=-\frac{r+r\cos^2(\gamma)+p\sin^2(\gamma)}{2}$.
The transformed payoff matrix is then-
\begin{equation}
U=\left(\begin{array}{c|cc} & C & Q \\\hline C & \frac{r-r\cos^2(\gamma)-p\sin^2(\gamma)}{2} & -\frac{r-r\cos^2(\gamma)-p\sin^2(\gamma)}{2} \\  Q & -\frac{r-r \cos^2(\gamma)}{2} & \frac{r-r \cos^2(\gamma)}{2}\end{array}\right).\nonumber
\end{equation} 
When compared to the Ising game matrix Eq.~(\ref{eq13-}), we get
$J+h=\frac{r-r\cos^2(\gamma)-p\sin^2(\gamma)}{2}$ and
$J-h=\frac{r-r\cos^2(\gamma)-p\sin^2(\gamma)}{2}$. Solving these simultaneous equations, we have $J=\frac{r-r\cos^2(\gamma)-p\sin^2(\gamma)}{2}$ and $h=0$. The magnetization which is the difference between the number of sites wherein quantum wins versus the number of sites wherein cooperation wins then is given by-
\begin{equation}
m=\frac{\sinh(\beta h)}{\sqrt{\sinh^2(\beta h)+e^{-4\beta J}}}=0.
\end{equation}
For quantum Prisoner's dilemma in the thermodynamic limit there is no unique Nash equilibrium wherein the sites can choose between cooperation and quantum strategy. The sites are equally divided between where cooperation wins or quantum strategy wins. 
\subsection{Quantum versus Defect}
Following on from the discussion in section~\ref{sec2.1}, we calculate the magnetization in a scenario when Alice and Bob have access to only quantum ($Q=iZ$) or defect ($D=X$) strategies. The payoff matrix is given by-
\begin{equation}\label{eq28}
U=\left(\begin{array}{c|cc} & Q & D \\\hline Q & r & t\sin^2(\gamma)+s\cos^2(\gamma) \\  D & t\cos^2(\gamma)+s\sin^2(\gamma) & p\end{array}\right).
\end{equation}
In this case the Nash equilibrium will change with $\gamma$. For example, when $\gamma=0$, the Nash equilibrium is to defect. However, when $\gamma=\pi/2$ the Nash equilibrium is the quantum strategy. Transforming the matrix using the method, explained in section~\ref{sec2}, we get the transformations as-
$\lambda=-\frac{a+c}{2}=-\frac{r+t\cos^2(\gamma)+s\sin^2(\gamma)}{2}$ and $\mu=-\frac{b+d}{2}=-\frac{p+t\sin^2(\gamma)+s\cos^2(\gamma)}{2}$. Thus, the transformed payoff matrix is-
\begin{equation}
U=\left(\begin{array}{c|cc} & Q & D \\\hline Q & \frac{r-t\cos^2(\gamma)-s\sin^2(\gamma)}{2} & \frac{t\sin^2(\gamma)+s\cos^2(\gamma)-p}{2} \\  D & -\frac{r-t\cos^2(\gamma)-s\sin^2(\gamma)}{2} & -\frac{t\sin^2(\gamma)+s\cos^2(\gamma)-p}{2}\end{array}\right).\nonumber
\end{equation} 
When compared to the Ising game matrix Eq.~(\ref{eq13-}), we get
$J+h=\frac{r-t\cos^2(\gamma)-s\sin^2(\gamma)}{2}$ and 
$J-h=\frac{p-t\sin^2(\gamma)-s\cos^2(\gamma)}{2}$. Solving these simultaneous equations, we have $J=\frac{r+p-t-s}{4}$ and  $h=\frac{r-p+(s-t)\cos(2\gamma)}{4}$ and the magnetization  for the N-site quantum prisoner's dilemma as in Fig.~\ref{fig8:} is, using Eq.~(\ref{eq13})-
\begin{eqnarray}\label{eq29}
m&=&\frac{\sinh(\beta h)}{\sqrt{\sinh^2(\beta h)+e^{-4\beta J}}}\nonumber\\
&=&\frac{\sinh(\beta \frac{r+(s-t)\cos(2\gamma)-p}{4})}{\sqrt{\sinh^2(\beta \frac{r+(s-t)\cos(2\gamma)-p}{4})+e^{-\beta (r+p-t-s)}}}.
\end{eqnarray} 
 The magnetization from Eq.~(\ref{eq29}) for $\gamma=0$ becomes-
\begin{eqnarray}
m=\frac{\sinh(\beta \frac{r+s-t-p}{4})}{\sqrt{\sinh^2(\beta \frac{r+s-t-p}{4})+e^{-\beta (r+p-t-s)}}},\nonumber
\end{eqnarray} 
which is same as derived in Ref.~\cite{3} for classical Prisoner's dilemma. We see that a phase transition for quantum Prisoner's dilemma occurs when-
\begin{eqnarray}\label{eq23}
& &\sinh(\beta \frac{r+(s-t)\cos(2\gamma)-p}{4})= 0,\nonumber\\
&\implies& r+(s-t)\cos(2\gamma)-p= 0\implies \cos(2\gamma)=\frac{r-p}{t-s}.\nonumber\\
\end{eqnarray}
\begin{figure}[t!]
\begin{center}
\includegraphics[width=.9\linewidth]{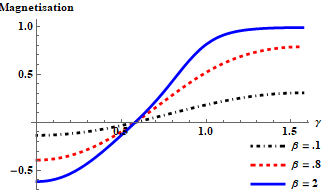}
  \caption{Magnetization versus $\gamma$ for quantum prisoner's dilemma taking $r=3$, $t=5$, $p=1$ and $s=0$ when players have access to quantum and defect strategy. A phase transition occurs at $\gamma=\frac{\cos^{-1}2/5}{2}=.579\  rad$.}
  \label{fig4:}
  \end{center}
\end{figure}
For $r=3$, $t=5$, $p=1$ and $s=0$, the phase transition from Eq.~(\ref{eq23}) should occur at $\gamma=\frac{\cos^{-1}(3-1)/(5-0)}{2}=.579\  rad$ as shown in Fig.~\ref{fig4:}. Also, it is quite interesting to note that from the magnetization in Eq.~(\ref{eq29}) for $\gamma=\pi/2$, the magnetization is always positive independent of payoffs of the payoff matrix Eq.~(\ref{eq15}) as for Prisoner's dilemma $r>p$. This implies that whatever is the value of reward, temptation or punishment, at a majority of the sites the quantum strategy wins. As we can see from Fig.~\ref{fig4:}, when $\beta$ increases or the temperature decreases, the number of players choosing quantum strategy increases in the region where magnetization is positive. However, in the regime where the magnetization is negative, the number of defectors increases when $\beta$ increases. The population tends to become unbiased towards both the choices when $\beta$ tends to $0$ even when there is a unique Nash equilibrium in the two player case. Comparing with the classical Prisoner's dilemma \cite{3} in the thermodynamic limit, we see that there is phase transition in the quantum case. This is interesting as in the thermodynamic limit for the classical Prisoner's dilemma majority were always defecting. However, we see here that for maximum entanglement, the defectors are always in  a minority. 
\section{Game of chicken}
The game of ``Chicken" refers to a situation where two players drive their bikes toward each other, each can either swerve or go straight\cite{4}. If the player swerves but the opponent doesn't, he can be called a coward or "Chicken". The payoff matrix for the game of chicken with $a=-s,\ b=r,\ c=-r$ and $d=0$ from Eq.~(\ref{eq9}) is given as-
\begin{equation}\label{eq18}
U=\left(\begin{array}{c|cc} & straight & swerve \\\hline straight & -s,-s & r,-r \\  swerve & -r,r & 0,0\end{array}\right),
\end{equation} 
where $``r"$ denotes reputation while $``s"$ denotes the  injury cost with the condition $s>r>0$. If one player drives straight and other swerves, the one who swerves  looses reputation, while other gains in reputation. However, a crash occurs injuring both if both drive straight at each other. There are two pure strategy Nash equilibriums (straight, swerve) and (swerve, straight) giving payoff $r$ to one player and -$r$ to other. There is a mixed strategy Nash equilibrium also which is given by $(\sigma,\sigma)$, where [$\sigma$ = $p$.straight+$(1-p)$.swerve] with $p=\frac{r}{s}$ ($p$ being the probability to choose straight).

Taking the classical payoff matrix as in Eq.~(\ref{eq18}), we calculate the quantum payoff's via the scheme as done in section 2.1. If both the players move straight, then it brings a higher loss to both the players. Thus straight strategy can be taken equivalent to  defection and swerve as cooperation. We thus have the classical strategies represented via the unitary matrices as- swerve$=I$, straight$=X$ while the quantum strategy is $Q=iZ$. The full  payoff matrix including both classical and the quantum strategies is therefore-
\begin{equation}\label{chick}
U=\left(\begin{array}{c|ccc} & swerve & straight & Q\\\hline swerve & 0,0 & -r,r  & \alpha 1,\alpha 1\\  straight & r,-r & -s,-s & \alpha 2,\alpha 3 \\Q & \alpha 1,\alpha 1 & \alpha 3,\alpha 2 &  0,0\end{array}\right),
\end{equation}
where $\alpha 1=-s\sin^2(\gamma)$, $\alpha 2=r\cos(2\gamma)$ and $\alpha 3=-r\cos(2\gamma)$. Now, we try to figure out what happens in the infinite player limit or the thermodynamic limit of quantum game of Chicken. To do this we break the quantum payoff matrix into two $2\times2$ blocks. This is what we do below, we first consider Quantum vs. Swerve and then Quantum vs. Straight.
\subsection{Quantum versus Swerve}
As discussed in section 2.1 we calculate the magnetization in a scenario when the qubits at a particular site are acted on by only quantum ($Q=iZ$) or  swerve ($I$) strategies. The payoff matrix is then given from Eq.~(\ref{chick}) for only row player as-
\begin{equation}
U=\left(\begin{array}{c|cc} & swerve & Q \\\hline swerve & 0 & -s \sin^2(\gamma) \\  Q & -s \sin^2(\gamma) & 0\end{array}\right).
\end{equation}
As we can see from the payoff matrix, there are two Nash equilibriums- both choosing swerve or both choosing quantum. To go to the thermodynamic limit of the two player quantum game of chicken we follow the same procedure as was done for two player quantum prisoner's dilemma. First transforming the matrix as given above using $\lambda=-\frac{a+c}{2}=\frac{s \sin^2(\gamma)}{2}$ and $\mu=-\frac{b+d}{2}=\frac{s \sin^2(\gamma)}{2}$, to map it to the Ising game matrix~(\ref{eq13-}), we have-
\begin{equation}
U=\left(\begin{array}{c|cc} & swerve & Q \\\hline swerve & \frac{s \sin^2(\gamma)}{2} & -\frac{s \sin^2(\gamma)}{2} \\  Q & -\frac{s \sin^2(\gamma)}{2} & \frac{s \sin^2(\gamma)}{2}\end{array}\right).\nonumber
\end{equation} 
When compared to the Ising game matrix Eq.~(\ref{eq13-}), we get $J+h=\frac{s \sin^2(\gamma)}{2}$ and $J-h=\frac{s \sin^2(\gamma)}{2}$. Solving these simultaneous equations, we have
$J=\frac{-s \sin^2(\gamma)}{2}$ and $h=0$. From Ising model, the magnetization is- 
\begin{equation}
m=\frac{\sinh(\beta h)}{\sqrt{\sinh^2(\beta h)+e^{-4\beta J}}}=0, \mbox{as $h=0$}.
\end{equation} 
Thus, the net magnetization vanishes in other words it is independent of $r,s,t,p$, i.e., the number of players choosing straight is exactly same as the number of players playing quantum in the thermodynamic limit.  Finally, we see that when $\gamma=0$, the quantum strategy $Q$ reduces to the swerve or cooperation strategy in the classical game of chicken. 
\subsection{Quantum versus Straight}
\begin{figure}[h!]
\begin{center}
\includegraphics[width=.9\linewidth]{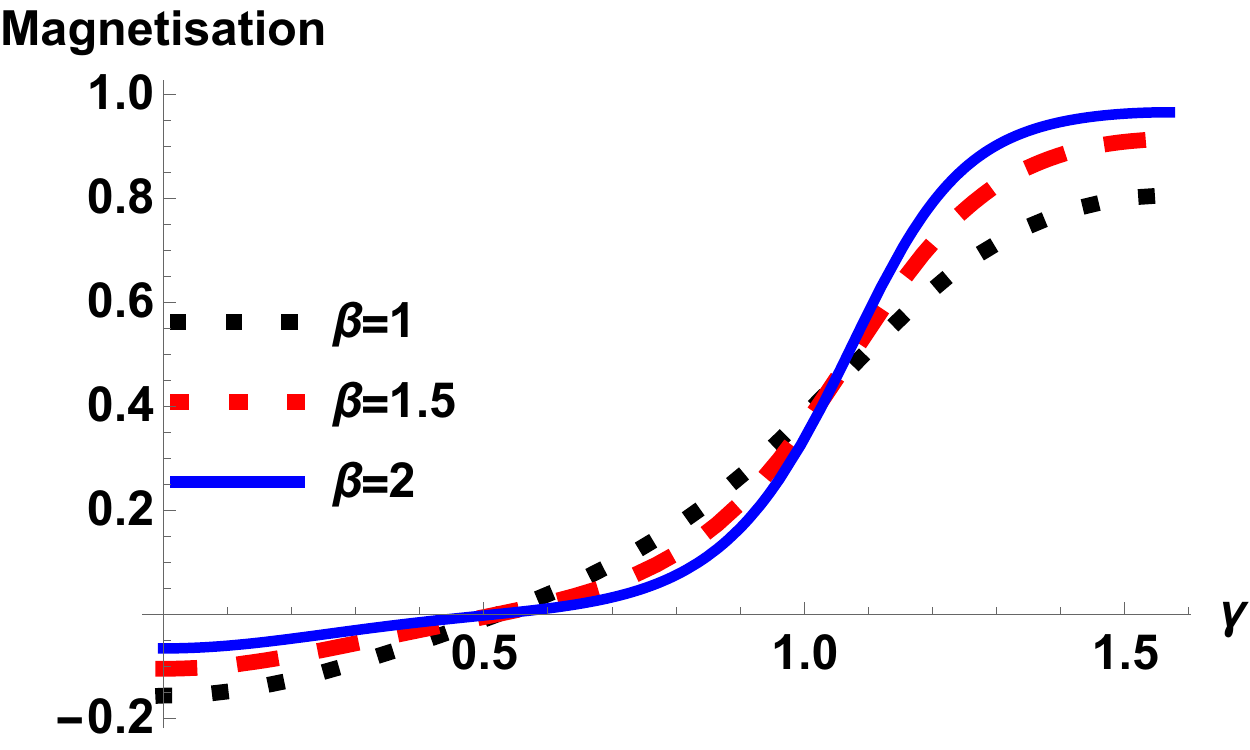}
  \caption{Magnetization versus $\gamma$ for quantum game of Chicken with $r=s=4$ for different $\beta's$ when players have access to quantum and straight strategy. For maximum entanglement i.e., $\gamma=\pi/2$, the magnetization is always positive irrespective of the values for $s,\ r$ and $\beta$.}
  \label{fig3:}
  \end{center}
\end{figure}
As discussed in section~\ref{sec2.1} we calculate the magnetization in a scenario when the qubits at a particular site are acted only by quantum ($Q=iZ$) or straight ($=X$) strategies. The payoff matrix is given by-
\begin{equation}\label{eq33}
U=\left(\begin{array}{c|cc} & Q & straight \\\hline Q & 0 & -r\cos(2\gamma) \\  straight & r\cos(2\gamma) & -s\end{array}\right).
\end{equation}
The Nash equilibrium in this case changes with $\gamma$. For $\gamma<\pi/4$, there are three Nash equilibria- (Quantum, Straight), (Straight, Quantum) and $(\sigma,\sigma)$, where $\sigma = p\times \mbox{quantum} +(1-p)\times \mbox{straight}$, with $p=\frac{s-r\cos(2\gamma)}{s}$. However, for $\gamma>\pi/4$, there is only one Nash equilibrium- both players choosing Quantum. Transforming the matrix~(\ref{eq33}) as given above using  $\lambda=-\frac{a+c}{2}=\frac{s+r\cos(2\gamma)}{2}$ and $\mu=-\frac{b+d}{2}=-\frac{r\cos(2\gamma)}{2}$, to map it into the Ising game matrix and thus calculate the Nash equilibrium in the thermodynamic limit, we have for the transformed payoff matrix-
\begin{equation}
U=\left(\begin{array}{c|cc} & Q & straight \\\hline Q & -\frac{r\cos(2\gamma)}{2} & \frac{s-r\cos(2\gamma)}{2} \\  straight & \frac{r\cos(2\gamma)}{2} & -\frac{s-r\cos(2\gamma)}{2}\end{array}\right).\nonumber
\end{equation} 
When compared to the Ising game matrix Eq.~(\ref{eq13-}), we get $J+h=-\frac{r\cos(2\gamma)}{2}$ and $J-h=\frac{r\cos(2\gamma)-s}{2}$. Solving these simultaneous equations, we have $J=-\frac{s}{4}$ and $h=\frac{s-2r\cos(2\gamma)}{4}$. Thus, the magnetization in the thermodynamic limit is-
\begin{equation}\label{eq35}
m=\frac{\sinh(\beta h)}{\sqrt{\sinh^2(\beta h)+e^{-4\beta J}}}=\frac{\sinh(\beta \frac{s-2r\cos(2\gamma)}{4})}{\sqrt{\sinh^2(\beta \frac{s-2r\cos(2\gamma)}{4})+e^{\beta s}}}.
\end{equation} 
The magnetization from Eq.~(\ref{eq35}) for $\gamma=0$ is then-
\begin{eqnarray}
m=\frac{\sinh(\beta \frac{s-2r}{4})}{\sqrt{\sinh^2(\beta \frac{s-2r}{4})+e^{\beta s}}},\nonumber
\end{eqnarray} 
which is the same as derived in Ref.~\cite{3}. The phase transition for the quantum game of chicken occurs when-
\begin{eqnarray}\label{eq30}
\sinh(\beta \frac{s-2r\cos(2\gamma)}{4})= 0&\implies& s-2r\cos(2\gamma)= 0\nonumber\\&\implies& \cos(2\gamma)=\frac{s}{2r}.
\end{eqnarray}
For $r=s=4$, the phase transition from Eq.~(\ref{eq30}) should occur at $\gamma=\frac{\cos^{-1}(1)/(2)}{2}=\pi/6$ as shown in  Fig.~\ref{fig3:}. It is to be noted  from the magnetization, see Eq.~(\ref{eq30}) for $\gamma>\pi/4$, the magnetization is always positive independent of payoffs of the payoff matrix Eq.~(\ref{eq18}) as for game of chicken $s>r>0$. This implies that independent of the reputation or injury cost, the majority of the population would always choose the quantum strategy. When fluctuation in choices become maximum or $\beta=0$, the players become unbiased towards quantum or straight even when a unique Nash equilibrium exists in the two player game.  
\section{Conclusions}
The aim in this work was to figure out the quantum Nash equilibrium in the thermodynamic limit. In the thermodynamic limit we see that, the quantum and cooperation strategy are equally probable. However, when the players have access to defection and quantum strategy, a phase transition occurs when the entanglement between players increases in favor of the quantum strategy. Further, when the entanglement is maximum then irrespective of payoffs, the majority always choose the quantum strategy and don't defect. Even in game of ``Chicken", the majority of players would always choose the quantum strategy over defection when the entanglement is maximum. Thus, we can conclude that when the players have access to quantum strategy, defection in a population reduces. {Further to point out the generality of our approach, instead of the quantization scheme of Eisert, et. al, see Ref.~[6], we could have chosen any quantization procedure of the two-player game and our approach could be used to extend the quantized game to the thermodynamic limit. An account of different quantization schemes of classical games can be found in Ref.~[10].}
\section{Acknowledgments}
This work was supported by the grant ``Non-local correlations in nanoscale systems: Role of decoherence, interactions, disorder and pairing symmetry'' from SCIENCE \& ENGINEERING RESEARCH BOARD, New Delhi, Government of India, Grant No.  EMR/20l5/001836,
Principal Investigator: Dr. Colin Benjamin, National Institute of Science Education and Research, Bhubaneswar, India. CB thanks Condensed matter and Statistical Physics section of the Abdus Salam ICTP, Trieste, Italy for funding a research visit during which a part of this work was completed.
\section{Author's contributions}
C.B. conceived the proposal, S.S. did the calculations on the advice of C.B., C.B. and S.S. analyzed the results and wrote the paper. Both authors reviewed the manuscript.
\section{Conflicts of Interest}
The authors have no potential financial or non-financial conflicts of interest.
\section{Data Availability Statement}
All data generated or analysed during this study are included in this manuscript.

\end{document}